# Performance Optimization and Parallelization of a Parabolic Equation Solver in Computational Ocean Acoustics on Modern Many-core Computer

Min Xu[1], Yongxian Wang[1,*], Anthony Theodore Chronopoulos[2,3], Hao Yue[1]

1 National University of Defense Technology, Changsha, Hu'nan 410073, China
2 University of Texas at San Antonio, San Antonio, TX 78249, USA
3 Visiting Faculty, Department of Computer Science, University of Patras, Greece
* Corresponding author. E-mail: yxwang@nudt.edu.cn

**ABSTRACT**

As one of open-source codes widely used in computational ocean acoustics, FOR3D can provide a very good estimate for underwater acoustic propagation. In this paper, we propose a performance optimization and parallelization to speed up the running of FOR3D. We utilized a variety of methods to enhance the entire performance, such as using a multi-threaded programming model to exploit the potential capability of the many-core node of high-performance computing (HPC) system, tuning compile options, using efficient tuned mathematical library and utilizing vectorization optimization instruction. In addition, we extended the application from single-frequency calculation to multi-frequency calculation successfully by using OpenMP+MPI hybrid programming techniques on the mainstream HPC platform. A detailed performance evaluation was performed and the results showed that the proposed parallelization obtained good accelerated effect of 25.77× when testing a typical three-dimensional medium-sized case on Tianhe-2 supercomputer. It also showed that the tuned parallel version has a weak-scalability. The speed of calculation of underwater sound field can be greatly improved by the strategy mentioned in this paper. The method used in this paper is not only applicable to other similar computing models in computational ocean acoustics but also a guideline of performance enhancement for scientific and engineering application running on modern many-core-computing platform.

**1. INTRODUCTION**

Sound is the only energy form that can transmit in long distance in seawater medium. It is the most effective method of target detection and information transmission underwater. [1] Using a three-dimensional numerical model it is difficult to provide real-time acoustic field data for sonar equipment and other applications. [2] Therefore, efficient large-scale parallel computing based on the latest high-performance computing platform can solve this problem.

Many scholars have published a lot of research studies about three-dimensional numerical underwater acoustic model. Based on these studies, it can be found that on the one hand the "weak three-dimensional approximation" approach, which completely ignores the coupling effect between different azimuths, is commonly used. [3]On the other hand, the parallel computing has not made full use of the current many-core computing power. Considering these problems, this paper proposes a parallelization and optimization combining the characteristics of architecture and application program to implement the efficient simulation of real three-dimensional underwater acoustic propagation problem on Tianhe-2 supercomputer.

## 2. NUMERICAL UNDERWATER ACOUSTIC MODEL

The Parabolic Equation method (PE) is a numerical model in underwater acoustics.[4] In this paper, we use FOR3D which is a representative application program of PE to obtain a parallel algorithm program for three-dimensional underwater acoustic propagation. FOR3D used in this article can be downloaded from the Online Marine Acoustic Library (http://oalib.hlsresearch.com). The entire acoustic field is shown in Fig. 1.

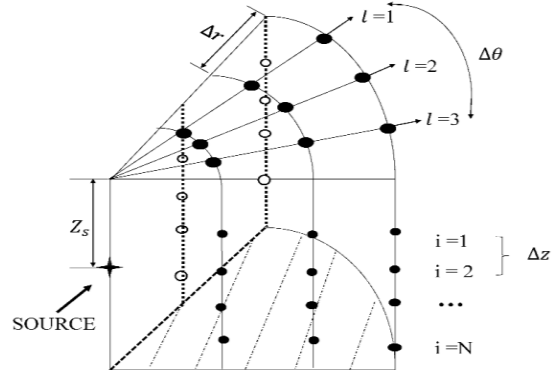

Fig. 1. The 3D view of one sector of acoustic field.

### 2.1 Mathematical Process of FOR3D Program

The starting point of parabolic Eq. (1) is the three-dimensional Helmholtz equation in cylindrical coordinates system $(r, \theta, z)$ as shown in Eq. (1):

$$\frac{\partial^2 p}{\partial r^2} + \frac{1}{r}\frac{\partial p}{\partial r} + \frac{1}{r^2}\frac{\partial^2 p}{\partial \theta^2} + \frac{\partial^2 p}{\partial z^2} + k_0^2 n^2 p = 0 \tag{1}$$

where $r, \theta, z$ denote range, azimuth and depth respectively, $p = p(r, \theta, z)$ denotes the pressure, $n = n(r, z)$ is the index of refraction and $k_0 = k_0(\omega)$ is the wave number.

By using Tappert's parabolic decomposition technique, [5] let $p(r, \theta, z) = u(r, \theta, z)w(r)$, where $w(r)$ is the Hankel function. Eq. (1) can further be expressed in terms of $u$ instead of $p$ as shown in Eq. (2) under the "wide angle assumption" proposed by Lee, Saad and Schultz .[6]

$$\frac{\partial^2 u}{\partial r^2} + 2ik_0 \frac{\partial u}{\partial r} + \frac{\partial^2 u}{\partial z^2} + \frac{1}{r^2}\frac{\partial^2 u}{\partial \theta^2} + k_0^2(n^2(r, \theta, z) - 1)u = 0 \tag{2}$$

By using the operator splitting and operator approximation technique and after a series of complicated calculations, we obtain Eq. (3) finally

$$\frac{\partial}{\partial r}u = \left(-ik_0 + ik_0\left[\left(1 + \frac{1}{2}x - \frac{1}{8}x^2\right) + \frac{1}{2}y\right]\right)u \tag{3}$$

where $x = k_0^2(n^2(r,\theta,z) - 1) + \frac{1}{k_0^2}\frac{\partial^2}{\partial z^2}$ and $y = \frac{1}{(k_0 r)^2}\frac{\partial^2}{\partial \theta^2}$ are two differential operators.[7]

To solve Eq. (3) numerically, a finite difference method is applied to discretize it in time and space. This results in an iteration-style of range advancing is formed as

$$\left[1 + \left(\frac{1}{4} - \frac{\delta}{4}\right)X\right]\left[1 - \frac{\delta}{4}Y\right]u^{j+1} = \left[1 + \left(\frac{1}{4} + \frac{\delta}{4}\right)X\right]\left[1 + \frac{\delta}{4}Y\right]u^j = RHS \qquad (4)$$

where $\delta = ik_0 \Delta r$, $X$ (and $Y$) is the difference version (operator) with respect with differential operator $x$ (and $y$). Superscript $j$ and $j+1$ indicate the current step and next step of range advancing respectively. To solve Eq. (4), the following two-step marching process is used: (a) Compute right-hand side ($RHS$) of Eq. (4) and obtain $v^{j+1}$ by solving Eq. (5). (b) Solve Eq. (6) and get the unknown $u$.

$$\left[1 + \left(\frac{1}{4} - \frac{\delta}{4}\right)x\right]v^{j+1} = RHS \qquad (5)$$

$$\left[1 - \frac{\delta}{4}y\right]u^{j+1} = v^{j+1} \qquad (6)$$

The distinct advantage of this two-step procedure is that only two tri-diagonal systems need to be solved for each step marched forward in the range. Consequently, less memory and fewer computations are required. Noticing that a central difference scheme introduced in operator $X$ and $Y$ will produce a tri-diagonal linear system for both Eq. (5) and Eq. (6). Without loss of generality, we denote the discretized form of both tri-diagonal systems as follows

$$A^{j+1}v_{m,l}^{j+1} = RHS \qquad (7)$$

$$B^{j+1}u_{m,l}^{j+1} = v_{m,l}^{j+1} \qquad (8)$$

where subscripts $l, m$ denote the index of grid in depth ($z$) and azimuth ($\theta$) direction respectively. Superscript $j$ denotes the index of the grid in the range direction and the grid is shown in Fig. 1.

## 2.2 Code and Algorithm of FOR3D

The main program of FOR3D consists of a series of subroutines. The main program contains the main loop to perform the calculation of the propulsive process by calling different subroutines to perform a variety of tasks as shown in Fig. 2.

## 3. OPTIMIZATION AND PARALLELIZATION OF THE THREE-DIMENSIONAL ACOUSTIC UNDERWATER MODEL

### 3.1 Optimization for FOR3D on a Single Node

3.1.1 Optimization options for compiling

At first, we compare the gfortran compiler with the ifort compiler. Then, we focus on the compiler options: `-O` (compiler automatically optimization), `-xHost` (generate instructions for the highest instruction set), `-ipo` (enable multi-file IP optimization between files), `-funroll-all-loops` (unroll loops), `-parallel` (enable the auto-parallelizer to generate multi-threaded code forloops that can be safely executed).

```
1  for each frequency ω of wideband signals
2    Initialization from the configurations of sound sources, receivers
     and sound speed profiles
3    set current range r = 0
4    while r <max_range
5      r = r + Δr
6      boundary condition
7      update matrix A and B in equation (7) and (8)
8      compute RHS in equation (4)
9      update unknown u by a tri-diagonal linear solver (two-step
       subroutine)
10     update pressure p = u * Hankel function
11   end while
12 next frequency
13 post-processing and output
```

Fig2. Algorithm of the parabolic-equation model used in FOR3D

3.1.2 Using efficient tuned mathematical library

In FOR3D, each stepping loop of the solution has to solve the linear equations of tri-diagonal matrix twice. The process can be replaced by the Math Kernel Library (MKL) developed by Intel to simplify the program structure.

3.1.3 Adjusting structure and order of the code

In order to fully reduce the memory requirements, the original program just reads the data required for the current calculation line by line, rather than completing the reading work only once. However, this operation restricts the further optimization for the process. We isolate the I/O operation and calculation with little sacrifice in memory in order to following optimization.

3.1.4 Vectorization

The Intel compiler provides automatic vectorization of the compiler options: `-vec-report`. In addition, we add the complier directive (`# pragma simd`) to the code of computing the tri-diagonal matrix for manual vectorization.

**3.2 Multi-node Optimization and Parallelization for FOR3D on Multiple CPU Cores**

3.2.1 Hotspot analysis of FOR3D

Using VTune Amplifier to analyze FOR3D and locate the hotspots. As shown in Fig.3, the hotspot of FOR3D is the subroutine twostep.

Fig3. The result of the hotspot analysis in VTune Amplifier.

3.2.2 Multi-threaded parallel technology

This paper mainly uses OpenMP scheme which is a shared storage to accomplish the parallel computing. [8] In the outermost loops of FOR3D, there exits data dependency between different azimuth and depths. But there is no dependency between azimuth and the depth in the same range mentioned in line 7 in Fig. 2 which can be accomplished by adopting task parallelism.[9] In addition, the loops in the computing subroutines mentioned in line 8 in Fig. 2 can be implemented by adopting data parallelism.[10] The parallel implementation of FOR3D is shown in Fig.4.The parts in the red dashed box are the main tasks for the OpenMP implementation.

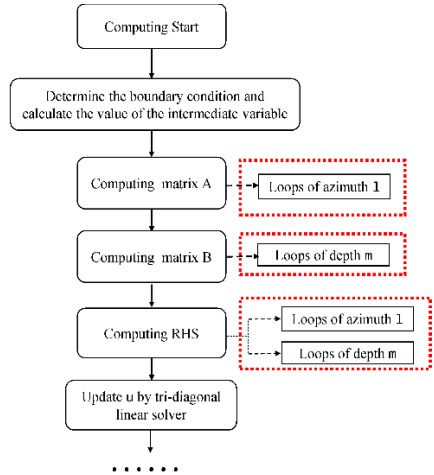

Fig4. The multi-thread parallel strategy of FOR3D

3.2.3 Multi-process optimization and parallelization of FOR3D

The original FOR3D can only calculate the propagation loss of a single-frequency acoustic source underwater in a single run. [11]In this paper, we proposed a parallel implementation for calculating the propagation loss simultaneously when the acoustic source rays multiple-frequency by using MPI based on distributed storage. [12]The parallel implementation steps are shown in Fig. 1. If we use the previous section of the OpenMP parallel computing for calculation of acoustic field in a specific frequency, we can achieve a hybrid MPI + OpenMP parallelization.

## 4. EXPERIMENTS AND ANALYSIS

### 4.1 Experiments Platform and Computer Environment

This section presents a numerical simulation of the three-dimensional acoustic propagation problem. The acoustic source frequency is 50Hz and its depth is 100m from the surface of the water. The depth of the water is 6000m. The cylindrical area of simulation takes the acoustic source as the center with radius of 1000m. So the size of space grids is 20000(Range)×9000(Azimuth)×6000 (Depth). The transmission loss in the case when $\theta= -180°$ is shown in Fig.5. The computer environment for the experiments is shown in Table I.

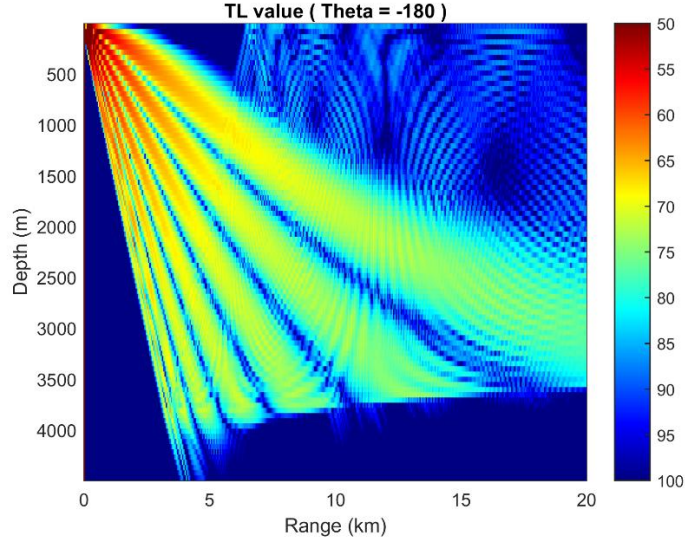

Fig.5. Transmission loss calculation in horizontally stratified medium using FOR3D

## 4.2 Results of Experiments

4.2.1 Serial optimization

A comparison of the results of the open source GNU Fortran compiler (gfortran) using the default option (-O0), with the commercial-grade Intel Fortran compiler (ifort) using the default option(-O2), is displayed in the first two rows in Table II. After changing the compiler, the performance of using the ifort compiler was improved more than 2 times. In the subsequent tests, we always use the ifort compiler. The result of selection of compiler options is also shown in Table II. Obviously, using -O3 -xHost –ipo   is better than the other options.

A comparison of the elapsed time of part code, which solves the tri-diagonal matrix   is shown in the first row in Table III. We found that the running time is reduced after our optimization. Using the performance-tuned math library MKL is 674s faster than the original code (about 11min).

The second row in TABLE III shows the elapsed time when running the entire program by using automatic vectorization and manual vectorization, respectively. In detail, the option -O2 is used when compiling the code optimized by manual vectorization. It can be seen that the effects of both vectorization schemes are very similar.

TABLE I. CONFIGURATION OF EXPERIMENTALHPCENVIRONMENT

| Name | Tianhe-2 Supercomputer |
|---|---|
| Node Configuration | 4 nodes×2CPU/node× 12cores/CPU× 2 hyper threads /core |
| CPU | Intel(R) Xeon(R) E5-2692 v2:2.20G Hz |
| Operating System | Kylin Linux |
| Compiler | GNU: gfortran(ver   4.4.7) |
| | Intel: ifort(ver 14.0.2) |

TABLE II. RESULT OF DIFFERENT COMPILING OPTIONS

| Options | Time(s) |
|---|---|
| -O0(gfortran) | 20400 |
| -O0(ifort) | 17160 |
| -O2 | 8602 |
| -O3 | 4643 |
| -xHost | 4415 |
| -ipo | 4577 |
| -funroll-all-loops | 5117 |
| -parallel | 7966 |
| -O3 -ipo -xHOST | 3705 |

TABLE III. TIME OF DIFFERENT OPTIMIZATION TESTS

| OptimizationMethods | | Time(s) |
|---|---|---|
| Using math library for tri-diagonal system | NoMKL | 2648 |
| | MKL | 1974 |
| Vectorization for range-stepping | No Vectorization | 8602 |
| | Automatic Vectorization | 4477 |
| | Manual Vectorization | 4305 |

4.2.2 Multi-threaded parallelization on a single node

    Solving the same problem serially on one node, the speedup of different number of threads is shown in Fig.6. As the number of threads increase, the time cost of the program decreases dramatically. When the number of threads exceeds 24, the running time increases. This is because multiple threads share the same processor core in this case, and multiple threads take turns using common hardware resources. Thus competition between the resources causes parallel speedup not to increase. In addition, too many threads' forking and joining operations also increase the time cost. The parallel efficiency reaches the maximum in four-threads running. On the other hand, using 16 threads can reach the maximum speedup. The reason why the parallel efficiency is not high may be explained by the fact that the problem size as well as the total amount of calculations of the simulation is not so large, and the system overhead and proportion of non-parallel region of code are both increasing when more threads are employed.

    The overall performance of combining the serial optimization with multi-threaded parallelization is reported in TABLE IV. The original serial program compiled by ifort with option−O2 is chosen as the base version. The result in TABLE IV shows the accelerating effects of different optimized versions. For the serial FOR3D program, the best performance on a single node with 24 cores gave a speedup of 25.77 compared with the base version.

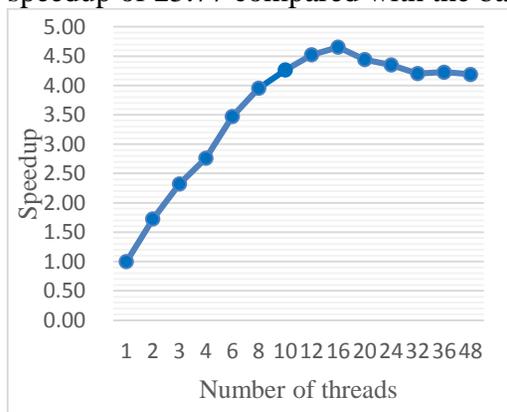 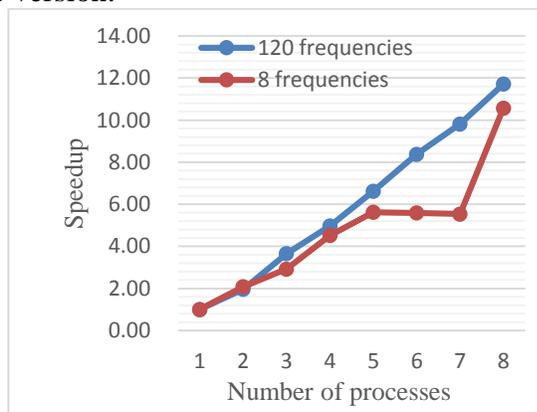

Fig.6. The performance of multi-threaded parallelization.

Fig.7. The performance of multiple MPI processes

TABLE IV. RESULT OF DIFFERENT OPTIMIZED VERSIONS OF FOR3D PROGRAM

| Versions | Base version (ifort –O2) | +Optimized Options | +MKL +Vectorization +Adjuststructure of code | +OpenMP |
|---|---|---|---|---|
| **Accelerating Effect** | 1 | 2.32 | 5.27 | 25.77 |

### 4.2.3 Multi-process hybrid parallelization

To evaluate the performance gain of parallelization method introduced in section 3.2.3, several numerical experiments are conducted on the HPC machine. The problem size of the test case, expressed in the number of grid points, is 2000 (distance) × 900 (azimuth) × 6000 (depth), and the wide band signals composed of both 8 frequencies and 120 frequencies are tested.

At first, we only use MPI multi-process parallelization and each process is mapped into a physical CPU core in the simulation. The performance result is shown in Fig 7. We can observe that the running time of the program decreases rapidly as the number of processes increase, and a good speedup can be obtained, especially in the 120-frequency case. However, in the 8-frequency case, there are only eight tasks to be assigned to multiple processes of which the number is varying from 1 to 8. This results in a low speedup due to severe lack of balance of the workloads among these processes. It is worthy to note that a super-linear speedup can be obtained in the case of 120 frequencies, which is due to the fact that multiple-process can make full use of the memory system of the single node. More data can be loaded into the memory all at once when using multiple processes, thus better data locality of the data access can be achieved. [13]

In the following test, the OpenMP multi-threaded and multi-process hybrid parallelization method are used, and the performance result is reported in Fig.8. The number of frequencies is 120, in these numerical simulation cases. It shows that different configurations of processes and threads have different performance results even if the number of total processor cores remains the same. We can also compare the accelerating effects between multi-process technique and multi-threaded technique. For an example, when taking the "3 MPI ranks (processes) × 4threads per rank" configuration ("3 × 4" shown in Fig. 8) as the baseline, doubling the number of MPI processes, i.e. "6 × 4" configuration, results in a parallel efficiency of 1.64, whereas doubling the number of threads per rank, i.e. "3 × 8" configuration, results in a parallel efficiency of 1.08. It's obvious that doubling the number of MPI processes has better performance than doubling the number of threads. The same conclusion can be drawn from other configurations.The main reason is that the multi-process implementation is a coarse-grain parallelization, which aims to partition the loads of different frequencies. By contrast, the multi-threaded implementation has fine-grain parallelism, which aims to accelerate the nested loops.

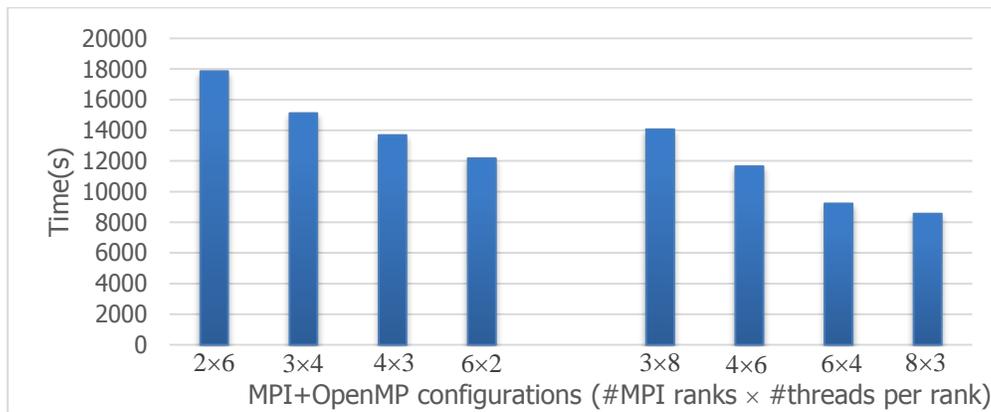

Fig.8. Performance of MPI+OpenMP hybrid parallelization for simulating the propagation of wide band underwater signals

## 5. CONCLUSION

Using high-performance computers is an effective way to improve the efficiency for calculations of the acoustic field. In this paper, we propose a series of parallelization and optimization methods to improve the performance of the three-dimensional parabolic equation solver and achieve good results. Based on the optimization results of single node, the multi-frequency parabolic equation solver implemented using OpenMP+MPI hybrid parallelization. The proposed parallelization and optimization methods can be easily extended to the distributed-memory environment composed of multiple nodes, and our work is helpful to the large-scale acoustic applications with real-time requirement.